MAX-PLANCK-INSTITUT FÜR EISENFORSCHUNG

**Max-Planck Institut für Eisenforschung GmbH**
Microstructure Physics and Alloy Design
Group leader: Dr. Baptiste Gault
Location: Max-Planck-Straße 1, 40237 Düsseldorf

Dr. Yue Li
Telephone: +49 211 6792 871
Telefax: +49 211 6792 333
Email: yue.li@mpie.de

# Convolutional neural network-assisted recognition of nanoscale $L1_2$ ordered structures in face-centred cubic alloys

Yue Li[1,*], Xuyang Zhou[1], Timoteo Colnaghi[2], Ye Wei[1],

Andreas Marek[2], Hongxiang Li[3], Stefan Bauer[4], Markus Rampp[2], Leigh T. Stephenson[1,*]

[1] Max-Planck Institut für Eisenforschung GmbH, Max-Planck-Straße 1, 40237 Düsseldorf, Germany

[2] Max Planck Computing and Data Facility, Gießenbachstraße 2, 85748 Garching, Germany

[3] State Key Laboratory for Advanced Metals and Materials, University of Science and Technology Beijing, 100083, Beijing, China

[4] Max Planck Institute for Intelligent Systems, Max-Planck-Ring 4, 72076 Tübingen, Germany

[*]Corresponding authors, Yue Li, Tel.: +49 211 6792 871, E-mail: yue.li@mpie.de

Leigh T. Stephenson, Tel.: +49 211 6792 794, E-mail: l.stephenson@mpie.de

**Abstract**

Nanoscale $L1_2$-type ordered structures are widely used in face-centred cubic (FCC) alloys to exploit their hardening capacity and thereby improve mechanical properties. These fine-scale particles are typically fully coherent with matrix with the same atomic configuration disregarding chemical species, which makes them challenging to be characterized. Spatial distribution maps (SDMs) are used to probe local order by interrogating the three-dimensional (3D) distribution of atoms within reconstructed atom probe tomography (APT) data. However, it is almost impossible to manually analyse the complete point cloud (>10 million) in search for the partial crystallographic information retained within the data. Here, we proposed an intelligent $L1_2$-ordered structure recognition method based on convolutional neural networks (CNNs). The SDMs of a simulated $L1_2$-ordered structure and the FCC matrix were firstly generated. These simulated images combined with a small amount of experimental data were used to train a CNN-based $L1_2$-ordered structure recognition model. Finally, the approach was successfully applied to reveal the 3D distribution of $L1_2$–type $\delta'$–$Al_3(LiMg)$ nanoparticles with an average radius of 2.54 nm in a FCC Al-Li-Mg system. The minimum radius of detectable nanodomain is even down to 5 Å. The proposed CNN-APT method is promising to be extended to recognize other nanoscale ordered structures and even more-challenging short-range ordered phenomena in the near future.

## Introduction

Traditionally, materials scientists investigate or characterize engineering materials by analysing a series of micrographs that reveal its complex microstructure at scales varying from the millimetre down to the nanometre. These are often manually done by individual scientists, sometimes aided by computational techniques [1,2]. These human-centred workflows have severe drawbacks, e.g. the demand for expertise, poor repeatability, and time-consuming processes. Since the development of materials informatics, further attention has been paid to improve the status quo through data-driven techniques, including machine learning and other artificial intelligence approaches [1-4]. As a core component of deep learning for image recognition, convolutional neural networks (CNNs) have the potential ability to speed up the analysis of micrographs and improve the repeatability of the analysis, and have the potential to reveal unforeseen patterns and details that would be hidden without application of advanced data-mining techniques [2,4]. In this work, we demonstrated the application of CNN for automatically recognizing nanoscale ordered structures within the point cloud of an atom probe reconstruction.

For crystalline materials, depending on the composition, preferential occupations of certain elements at specific crystallographic sites can take place, resulting in the occurrence of ordered nanoparticles in alloys. For face-centred cubic (FCC) alloys, the existence of a high-density dispersion of coherent $L1_2$ ordered nanoparticles have been highly desirable to enhance their mechanical performance, such as $Al_3Sc$ particles in additive manufacturing aluminium alloys [5,6], $\gamma'$ precipitates in superalloys [6-8], and $L1_2$-type nanodomains in some high-entropy alloys [9-11]. Currently, it is a hot topic to accurately recognize these nanodomains with the help of developing advanced characterization tools or analysis algorithms and further reveal an alloy's structure-property relationship.

There are two mainstreams to handle this task: transmission electron microscopy (TEM) and atom probe tomography (APT). TEM can capture accurate crystallographic information, even down to the atomic scale, but it is not generally a three dimensional (3D) technique and TEM diffraction patterns correspond to the interaction volumes through which the electron beam pass resulting in unspecific results. 3D information is generally not available in such studies and atomic mass information is also lacking [12,13]. As a comparison, APT is capable of detecting 3D elemental distributions with a near-atomic spatial resolution (the resolution of the best scenario is 0.3 nm in the lateral direction and 0.1 nm in the depth direction) and high chemical sensitivity (10–100 ppm) [12,14,15]. However, two significant drawbacks limit its application to precisely reveal the crystallographic and chemical information of fine-size ordered structures: the detection efficiency, limited to 50–80%, as well as trajectory aberrations and associated reconstruction artefacts [15]. Although a single atom cannot be imaged precisely by APT, a large number of reconstructed field evaporated atoms can provide statistical distributions of different elements and local crystallographic information can be revealed by spatial distribution maps (SDMs) [15,16]. SDMs are produced by first calculating the 3D offset vectors between each atom in either a local or global set. These offsets are then accumulated into 3D voxelized histograms, 2D histograms (projected upon a plane, commonly the *xy*, *yz*, or *zx* plane), or 1D histograms (commonly along the *z*-axis). In this work, only the 2D *zx*-SDM and the 1D *z*-SDM were used. Previously, SDMs have been applied to investigate ordered structures in certain binary [14,16] and ternary alloys [17,18]. Different structures generate distinctly different SDMs signatures and thus desired ordered structures can be recognized within APT data for further revealing the structure-property relationship. However, to analyse large APT datasets and recognize a large number of SDMs images (about 100 thousands of images in this study) is a computationally intensive task and almost impossible if manually attempted. Thus, an automatic ordered structure identification method is required. Note that the previous solution [17,18] is to exploit the difference in compositions between ordered structures and matrix and then analyse SDMs of different subsets divided by isocomposition surfaces of a certain composition, like at 8 at.% Li in an Al-Li-Mg system [17]. However, the filtering method cannot ensure that atoms from the matrix are not included, i.e., it is a little arbitrary to choose 8 at.% Li as the dividing line [17]. Moreover, this approach will become invalid when there is little difference in compositions between ordered structures and matrix, like the $L1_0$ ordered structure in Au-43Cu-7Ag (at.%) alloy [19], or when characterizing short-range ordered structures [13,20,21].

Recently, machine learning has been applied to APT data to automate the identification of a specimen's crystallographic orientation or improve microstructural feature extraction [22,23]. Machine learning algorithms have the potential to unveil ordered structures by learning characteristic patterns in experimentally obtained SDMs. As a representative in the field of image recognition, CNNs have been used to automate the identification of microstructural and crystallographic features using micrographs [4,24,25]. A remarkable advantage of CNNs is the automatic extraction of features with minimal human intervention [26,27]. The essence of image recognition via CNNs is to extract different levels of features such as low-level edges and colour features as well as more abstract features through a series of convolutional and pooling layers. Different crystal structures will generate different interplanar spacing in *zx*-SDM patterns with different relative colour scaling. These edges and colour features in SDMs are the key pieces of information for determining the structural types. Another advantage of CNNs for computer vision is its translation invariance but this phenomenon was not met in our study due to the pre-defined image generation procedure.

In this work, a CNN-based strategy is proposed to automatically recognize nanoscale $L1_2$-type ordered structures in FCC-based alloys using APT data with an ultra-high recognition ability. Firstly, a crystal structure library was built to include a wide range of possible configurations to then feed into producing many simulations of APT data, all based on either the $L1_2$ or FCC crystal structure. From these simulated structures, the corresponding *zx*-SDMs along with specific crystallographic direction were generated. The obtained SDMs (used as inputs) combined with their corresponding crystal structures (used as labels) were divided into training, validation, and test datasets, which were then used to train a CNN to generate an $L1_2$ ordered structure recognition model. A second training procedure was also performed after enriching these synthetic datasets with few experimentally obtained data, to enhance the model performances and speed up the training. Finally, the experimentally-obtained SDMs from an Al-Li-Mg alloy were input into this recognition model to identify the 3D distributions of the $L1_2$–type $\delta'$–$Al_3$(LiMg) particles in the FCC matrix. This result was further compared with the previous isocomposition approach to highlight its advantages.

## Results

### APT data analysis

The typical 3D atom probe tomography of Al-6.79Li-5.18Mg (at.%) alloy is shown in Figure 1 (a). Four crystallographic poles are observed from Figure 1 (b), corresponding to $[\bar{1}01]$, $[\bar{1}11]$, $[\bar{1}02]$, $[\bar{1}13]$, respectively. The 3D atom probe tomography along $[\bar{1}01]$ crystallographic pole is reconstructed and presented in Figure 1 (c). Figure 1 (d) provides the close-up of a thin slice in Figure 1 (c), where atomic planes are imaged clearly, suggesting the high resolution in the depth direction which is necessary for generating the *zx*-SDMs and the *z*-SDMs. The isocomposition surfaces containing more than 8 at.% Li are visualized.

A parallel python-based program code was made to quickly scan the dataset shown in Figure 1 (c) via a cube with the side length of $a$ and generate a large number of the corresponding *zx*-SDMs along $[\bar{1}01]$. Figure 1 (e) and (f) gives two examples of *zx*-SDMs corresponding to FCC matrix and $L1_2$ precipitate voxels ($a$ = 4 nm), respectively. Note that only the *zx*-SDMs of Al-Al pairs are shown because the *zx*-SDMs of other element pairs cannot provide useful information due to the limited amount of data in these voxels. The interplanar spacing as shown in Figure 1 (f) is twice as large as the one in Figure 1 (e). This difference allows one to recognize the structural difference and thus highlight the sites of $L1_2$ particles.

Simulation of *zx*-SDMs

For the machine learning part, the first step was to build the synthetic dataset to train CNN parameters. Here, a crystal structure library was built to include various possible configurations, around the FCC and $L1_2$ crystal structures. Several parameters were considered to construct the crystal structure library, including lattice parameters, crystallographic rotation, simulating limited spatial resolution, loss of atoms simulating imperfect detection efficiency, and data size (number of atoms). Then, the simulated *zx*-SDMs along with $[011]$ were generated from these simulated configurations. Note that $[011]$ is equivalent to the $[\bar{1}01]$ pole shown in Figure 1 (c).

A procedure to generate the simulated *zx*-SDMs is shown in Figure 2. Firstly, the FCC and $L1_2$ crystal structures with a volume of $25\text{nm}^3$ were built based on the posgen program [28], as shown in Figure 2 (a). The lattice constant of the FCC-Al structure was defined as 0.405 nm [29,30]. The same lattice constant was set for the $L1_2$-type $Al_3(MgLi)$ structure due to the fully coherent nature [17]. Note that Mg and Li atoms were not separately labelled in this model because only the *zx*-SDMs of Al-Al pairs were used to make structure recognition in this paper. Then, the Euler transformation was applied to change the projection pole from $[001]$ to $[011]$, as shown in Figure 2 (b). Thirdly, certain levels of Gaussian noise were added to shift the atoms in x, y, and z reconstruction directions to model finite spatial resolution, as shown in Figure 2 (c). Note that the standard deviation (σ) of Gaussian noise in the z-direction is smaller than those in x and y directions, simulating the higher resolution in the depth direction. Fourthly, a certain fraction of atoms were randomly removed to simulate imperfect detection efficiency (see Figure 2 (d)). Finally, the corresponding *zx*-SDMs of Al-Al pairs of two crystal structures with different parameters were generated to build the *zx*-SDMs dictionary, as shown in Figure 2 (e).

Table 1 summarizes the parameters used for building a crystal structure library and generating the corresponding *zx*-SDMs dictionary. Two kinds of crystal structures were included with different noise levels and detection efficiencies. The upper and lower boundaries of the Gaussian noise levels were chosen based on the similarity between the simulated and measured *zx*-SDMs. The adjusted detection efficiency is used to change the number of atoms, and thus it is not kept a fixed experimental value. The generated *zx*-SDMs dictionary was augmented by generating more *zx*-SDMs which were rotated certain angles randomly. The rotation augmentation was applied to simulate the observed small-angle pattern rotations in the experimental *zx*-SDMs. In total, 18416 simulated images were included in this dictionary and divided almost equally into two classes. A noise estimation [31] was made on the simulated and experimental images, respectively, as shown in Supplementary Figure 1 (a) and (b). The simulated data can well embrace the noise level of the experimental *zx*-SDMs. Note that adding different levels of noise to the input data of the CNN can help in out-of-distribution generalization and transfer to the real data [32].

Network configuration

The simulated z-SDMs were split into 90% for training and validation and 10% for test. Five-fold cross validation was exploited to train the model. The used images were 150×150 pixels greyscale images with one channel of input, whose pixel values were between 0 (black) and 255 (white). The adopted CNN is shown in Figure 3 (a) and it consists of a six-layer structure with (plus four convolutional max pooling) layers and two fully connected layers (containing the last output layer). The detailed architecture of each layer is shown in Figure 3 (b). Note that other CNN configurations have been tested but exhibited quite high loss values of training and validation as shown in Figure 3 (c).

Two cases were performed: (1) all training and validation datasets consist of simulated data while test dataset consists of simulated data and experimental data; (2) training, validation, and test datasets consist of simulated data and experimental data. Here, 7 epochs were used to find the minimum cross-entropy loss and the entire training process only took approximately 14 minutes on an Intel Core i7-9700 CPU 3.00GHz.

Training, validation and test results

For case 1, Figure 3 (c) shows the final loss values of training and validation of the optimized CNN after 7 epochs. The history of loss values is shown in Supplementary Figure 2 (a). This model trained only on synthetic data reaches a 100% classification accuracy on the test data, which suggests that no overfitting occurs. Thus, L2 regularization was not further applied in this CNN. Nevertheless, it does not generalize to real experimental data, yielding a poor classification accuracy on a small test dataset composed of 48 experimental *zx*-SDMs of the Al-Li-Mg system, as shown in Figure 4 (a). It was found that several images corresponding to the FCC structure were wrongly given high $L1_2$ probabilities, like image number 0 and 47. This is mainly because the simulated *zx*-SDMs still cannot fully imitate the complex experimentally-obtained *zx*-SDMs of the FCC structure.

To solve this, in case 2, only 12 experimental *zx*-SDMs corresponding to the FCC structure were augmented into 84 samples using the same method as the simulated data and added to train CNN. The same procedure was implemented and a model was obtained with similar training and validation losses of about $1.3 \times 10^{-4}$. The history of loss values is shown in Supplementary Figure 2 (b). This model was also tested using the 10% test dataset (only containing the simulated data) and 100% classification accuracy was made. The predicted results of the 48 experimental *zx*-SDMs were shown in Figure 4 (b). As compared with Figure 4 (a), this model exhibited very good prediction ability on both the simulated and experimental data. When a near-zero value was predicted, the test image is close to the *zx*-SDM of FCC structure, while a near-one value signified similarity with the *zx*-SDM of $L1_2$ structure. Note that a value around 0.5 means that the test image is a mix of the two structures, as shown the image number 27 in Figure 4 (b), or other invalid images due to limited data (Supplementary Figure 3).

To more quantitatively illustrate the model's performance, a receiver operating characteristic curve (ROC) analysis [33,34] was further made on the 48 experimental images for the two cases, as shown in Figure 4 (c). The area under ROC curve (AUC) is 0.995 in case 2, which is higher than 0.963 in case 1. The higher AUC value suggests better model classification ability.

Application to Al-Li-Mg APT data

After verifying the obtained ordered structure recognition model in a small test dataset, it was further applied to the big dataset shown in Figure 1 (c). The side length $a$ of the scanning cube was set as 4 nm, and the scanning stride was 1 nm. Each smaller cube with a 1-nm side length was divided and its probability was represented by the sum of the predicted probabilities of all overlapped 4-nm cubes. Thus, the range of the summing probabilities was between 0 and 64. The frequency distribution of the $L1_2$ structure probabilities of 98175 4-nm cubes is shown in Supplementary Figure 3. Similar to Figure 4 (b), the lower value is closer to FCC while the higher value is to $L1_2$ structure. Figure 5 (a) shows the frequency distribution of the $L1_2$ structure probabilities of the 1-nm voxels. Two distinct peaks were observed at close to the values of 0 and 64, which corresponded to the FCC matrix and $L1_2$ structure, respectively. The data from three zones in Figure 5 (a) are extracted and their corresponding *zx*-SDMs are illustrated in Figure 5 (b). The *zx*-SDM of zone 1 exhibits obvious $L1_2$ signature, while this signature is unclear in zone 2 or 3. Finally, the value of 62 was taken as separating the two different crystal structures, which is closer to the AUC value in case 2 multiply by 64, i.e., 63.68. Figure 6 (a) shows the 1-nm voxels map with the $L1_2$ structure probability above 62. The corresponding nanoparticle size distribution is shown in Supplementary Figure 4, but more APT data is needed to give a statistical result. This will be used as an input into microstructure and strength models to build the structure-property relationship [30]. Note that the recognized minimum precipitate radius can be down to 0.5 nm, suggesting the ultra-high recognition ability. Figure 6 (b) and (c) show the species-specific z-SDMs for the segmented FCC matrix and $L1_2$ structures, respectively, plotted with arbitrary units for ease of comparison. All peak-peak distances in Figure 6 (c) corresponded to the interplanar spacing of the FCC matrix, while all peak-peak distances in Figure 6 (b) were twice as those in Figure 6 (c). As a comparison, Figure 6 (d) shows the previous atom map obtained using above 8 at.% Li isocomposition [17]. The corresponding species-specific intensity distributions along the *z*-axis in the *z*-SDMs of above and below 8 at.% Li are shown in Figure 6 (e) and (f), respectively, showing the same periodicities as seen in Figure 6 (b) and (c). Moreover, the sites of precipitates from the two methods are very similar.

Visualization of the CNN model

Deep learning models are often treated as black box methods. To understand where the obtained CNN model is looking in an image, the authors employed two kinds of visualization methods: feature maps and gradient-weighted class activation mapping (Grad-CAM). Feature maps are the outputs of the convolutional and pooling layers after applying different filters to the input image, which helps us learn how the proposed layer structure processes an input image [35,36]. Figure 7 (a) exhibits the feature maps learned by the first and fourth convolutional layers from two different *zx*-SDMs corresponding to FCC and $L1_2$ structures. After applying the filters in the first convolutional layer, a lot of versions of the FCC or $L1_2$ *zx*-SDMs were portrayed with different features highlighted in Figure 7 (a). For example, some focus on the edges, while others highlight the foreground or background. As going deeper into the CNN structure, the model identifies more abstract concepts. At this step, we often cannot interpret these deeper feature maps. In deeper layers, the model identifies more abstract concepts.

The output of Grad-CAM is a heatmap visualization for a given class label [37,38]. We can use this heatmap to visually verify where in the image the CNN is looking. As can be seen from Figure 7 (b), the obtained CNN model mostly focuses on the interplanar spacing feature in the deeper convolutional layers, which is the desired result.

## Discussion

In this paper, a CNN-assisted APT approach has been successfully applied to recognize $L1_2$ ordered precipitates in the FCC matrix with an ultra-high recognition ability. The proposed CNN-APT approach has several advantages over the traditional method based on isocomposition thresholding. The most important is that the traditional method is only based on the differences in compositions, while the present method attempts to take into account the entire crystal structure information including the occupancy sites and types of different atoms, more exactly, how this crystallographic information manifests its signature in 2D *zx*-SDM images. This enables the proposed method to precisely recognize ordered structures in different crystal materials. In terms of the traditional method, on one hand, it is arbitrary to choose one value to filter matrix data based on the isocomposition, like below 8 at.% Li in this Al-Li-Mg system (Figure 6 (d)). It is hard to ensure that matrix atoms are not included in such precipitate characterisation, which will significantly affect the composition measurements [17]. The proposed CNN approach has the capability of classifying the different crystal structures distinguishably. As shown in Figure 5, two obvious peaks were observed and 62 was reasonably chosen to filter the data. This chosen threshold matched well with the obtained AUC value in case 2 multiply by 64, i.e., 63.68. As shown in Figure 6, the average radius of precipitates from the isocomposition and proposed methods is 2.59 ± 0.9 and 2.54 ± 1.03 nm, respectively. On the other hand, the isocomposition method could fail when encountering the weak differences in compositions, especially like short-range ordered structures occurring in Ti and high-entropy systems [13,20,21,39]. The proposed method based on full crystal structure information is quite promising to handle this challenge in the near future.

Moreover, a 1-nm voxel, i.e., with 0.5 nm radius, is identified as the $L1_2$ structure only when the sum of the predicted probabilities of the surrounding overlapped 4-nm voxels is above 62. This suggests that the average predicted probability of the individual 4-nm voxel is above 0.96875. As shown in Supplementary Figure 3, the average value corresponds to the clear $L1_2$ structure signature in the 4-nm voxel. Thus, this ensures that the 1-nm voxel most likely is the $L1_2$ structure. The supplementary Figure 5 (a) and (b) gives an example of the 2D *zx*-SDM and 1D Al-Al *z*-SDM of a 1-nm voxel extracted from Figure 6 (a). This shows a slight $L1_2$ signature. However, this signal is weak and that is why the authors finally employed larger voxels to make the CNN recognition. Overall, the minimum radius of the detectable nanoparticle is down to 0.5 nm using the proposed method.

The CNN employed in this work handles a piece of zone of one image using filters, which enables the neural networks to watch a field rather than a pixel [24]. Each convolutional layer contains several filters to scan this image using a specific size kernel. As shown in Figure 7, through the first convolutional layer, clean edge and grey features are detected. With the deepened convolutional layers, more abstract and sparser features are obtained, and the CNN model mostly focuses on the desired interplanar spacing feature. If the deeper convolutional layers are not added, a higher loss value will be obtained like Figure 3 (c). This ablation study highlights the importance of deeper convolutional layers. Two case studies have been performed to train the CNN with or without real experimental data. The obtained models from the two cases exhibited quite a high classification accuracy in the simulated test dataset, but only the second case with some experimental data performed well in both simulated and experimental test datasets. This is attributed to the more complex situation occurring in experimental data which has not been fully involved in the simulated data.

It is worth mentioning that the transfer learning was considered in the beginning but finally given up. Firstly, applying the transfer learning requires enough data to train the neural networks [40]. Here, we only employed 12 experimental *zx*-SDMs corresponding to the FCC structure which were further augmented into 84 images. These should not be enough for performing the transfer learning. Moreover, as mentioned above, the addition of noise helps in out-of-distribution generalization and transfer to the real data [32]. In the case 1 with only simulated data, the poor performance is mainly that several images corresponding to the FCC structure were wrongly given high $L1_2$ probabilities. Thus, the authors utilized a small set of experimental data to further embrace the complex FCC structure. This made the present model perform well in making the $L1_2$ structure recognition.

In this paper, *zx*-SDMs have been successfully used to make $L1_2$ structure recognition via CNN. Another possible analysis way is to deal with the experimental *z*-SDMs (like Figure 6 (b) and (c)) and some curve analysis methods could be performed, like 1D CNN. In addition, the potential of applying 3D CNN to directly handle 3D APT cloud points could be explored, although it may be quite difficult. Note that the uncertainty quantification [41,42], as a non-trivial question and current research hot topic [43], should be considered in the next step, which is one of the challenges of using CNNs for scientific application. Moreover, the proposed method can easily be extended to other ordered structures encountered in FCC alloys. One only needs to build an appropriate crystal structure library and corresponding *zx*-SDMs dictionary. The methodology could also be extended to BCC and HCP alloying systems in the future. It should be pointed out that the success of the proposed CNN-APT method requires the occurrence of the pole structures in the detector event histogram (like Figure 1 (b)) where APT data exhibits the high depth resolution. The tomographic reconstruction is often calibrated by using the pole structures [17]. The pole information can be found in various metallic materials, such as aluminium, magnesium, and titanium alloys. There are several other methods to extract precipitates within matrix using APT data, such as pair correlation function (PCF) and K-nearest neighbour (KNN) distance analysis [15]. Two obvious advantages of the two approaches are that they do not require the occurrence of pole structure and involve examining the average local neighbourhood as a function of distance along all directions. A drawback is that the information along the lateral direction having the lower resolution could hinder the recognition of small-scale clusters/precipitates. In fact, the SDM technique is very similar to the PCF but along a particular crystallographic direction with the higher resolution, and thus the best structure information can be exploited to reveal the potential clusters/precipitates. In the future, it is promising to explore to make ordered structure or cluster recognition by coupling the proposed CNN framework with those other approaches based on localized compositional measurements or solute pair distances, especially when no pole occurs.

In conclusion, this is a demonstration of the potential of the CNN-based method for ordered structure recognition within APT data. It is demonstrated that this image recognition approach has the capability of revealing nanoscale $L1_2$ ordered particles in FCC system using simulated *zx*-SDMs and a small amount of experimental data. The minimum radius of detectable nanodomain can be down to 0.5 nm. As compared to the traditional method based on isocomposition, the proposed CNN-APT approach is more outstanding in revealing $L1_2$ ordered precipitates with the average radius of 2.54 nm in the FCC Al-Li-Mg system. The next work is to extend this proposed methodology to more challenging short-range ordering phenomena.

## Methods

APT experiments

The studied APT data (45 million) of Al-6.79Li-5.18Mg (at.%) (Al-1.8Li-5Mg (wt.%)) aged for 8 h at 423 K is from Ref. [17]. The Cameca Inc. LEAP 3000XSi was applied to gather atomic-scale data with a 55% detection efficiency. IVAS 3.8.4 was used to make data reconstruction and visualization. The reconstruction parameters, i.e., the field factor and image compression factor, were calibrated by the method introduced in Refs. [44,45].

Convolutional neural networks

For the used CNN, all layers used ReLu (Eq. 1) as the activation function, except the output layer which used Sigmoid (Eq. 2) for classification purposes.

$$\mathrm{ReLu}(x) = \max(0, x) \tag{1}$$

$$\mathrm{Sigmoid}(x) = 1/(1 + exp(-x)) \tag{2}$$

The two kinds of activation functions were applied to each neuron in CNN to determine whether the neuron should be activated or not. They also help normalize the output of each neuron to a range between -1 and 1 or between 0 and 1. The binary cross-entropy [46] was chosen as the loss function, which is often used to train a binary classifier. The loss value is given:

$$\text{Loss} = -\frac{1}{N}\sum_{i=1}^{N}(y_i.\log(p(y_i)) + (1 - y_i).\log(1 - p(y_i))) \quad (3)$$

where $y$ is the label (0 for FCC and 1 for $L1_2$ structure) and $p(y)$ is the predicted probability of each image corresponding to the $L1_2$ structure for all $N$ images. The training was performed using RMSprop optimization [47] with certain learning rate. RMSprop is an optimization method used to iteratively update the parameters in the CNNs so that the loss value is minimized. Different learning rates have been tested and the optimum value of 0.001 was finally chosen in this study. Here, the training and validation datasets were divided into several mini-batches with the size of 32, i.e., images were randomly chosen without replacement per epoch. For neural networks, several iterations are performed to train the entire dataset once which is called one epoch. The CNN was implemented using Keras 2.2.4 [48] with the TensorFlow 1.13.1 backend [49] on Python 3.7.

ROC curve

The ROC curve is applied to illustrate binary-classifier performance [33,34]. The curve is created on two basic evaluation measurements: true positive rate and false positive rate. The former is a performance measurement of the positive part, while the latter is of the negative part. The CNN model provides a probability score of $L1_2$ structure for each tested image. Model evaluation measures are calculated by moving threshold values across the scores. The AUC is calculated for model comparison.

## Data availability

The key data that support the findings within this paper can be found at the GitHub address https://github.com/a356617605, and other data are available from the corresponding authors upon reasonable request.

## Code availability

The codes about the simulation of *zx*-SDMs and CNNs model are available at the GitHub address https://github.com/a356617605, and other codes are available from the corresponding authors upon reasonable request.

## Acknowledgements

We acknowledge support from the Max Planck research network on big-data-driven matrials science (BiGmax). Luka Stanisic and Markus Kühbach are acknowledged for their help with writing codes. Baptiste Gault, Isabelle Mouton, Chang Liu, Ning Wang, Ziyuan Rao, Dengshan Zhou, and Qiang Du are acknowledged for fruitful discussions. The experimental data was acquired at the Australian Microscopy & Microanalysis Research Facility (AMMRF) at The University of Sydney. The material was provided by Christophe Sigli and Alexis Deschamps.

## Competing interests

The authors declare no competing interests.

## Author contributions

Y. Li and L.T. Stephenson planned the study. Y. Li made APT data analysis, implemented the simulation of *zx*-SDMs, built CNNs, and applied this obtained recognition model to Al-Li-Mg APT data. L.T. Stephenson and T. Colnaghi further optimized these codes. All authors discussed the results and were involved in the writing of the manuscript.

## References

1 Oviedo, F. *et al.* Fast and interpretable classification of small X-ray diffraction datasets using data augmentation and deep neural networks. *npj Comput. Mater.* **5**, 60, (2019).

2 Aguiar, J. A., Gong, M. L., Unocic, R. R., Tasdizen, T. & Miller, B. D. Decoding crystallography from high-resolution electron imaging and diffraction datasets with deep learning. *Sci. Adv.* **5**, eaaw1949, (2019).

3 Butler, K. T., Davies, D. W., Cartwright, H., Isayev, O. & Walsh, A. Machine learning for molecular and materials science. *Nature* **559**, 547-555, (2018).

4 Shen, Y.-F., Pokharel, R., Nizolek, T. J., Kumar, A. & Lookman, T. Convolutional neural network-based method for real-time orientation indexing of measured electron backscatter diffraction patterns. *Acta Mater.* **170**, 118-131, (2019).

5 Li, R. *et al.* Developing a high-strength Al-Mg-Si-Sc-Zr alloy for selective laser melting: Crack-inhibiting and multiple strengthening mechanisms. *Acta Mater.* **193**, 83-98, (2020).

6 Deschamps, A. *et al.* Experimental and modelling assessment of precipitation kinetics in an Al–Li–Mg alloy. *Acta Mater.* **60**, 1917-1928, (2012).

7 Tu, Y., Mao, Z. & Seidman, D. N. Phase-partitioning and site-substitution patterns of molybdenum in a model Ni-Al-Mo superalloy: An atom-probe tomographic and first-principles study. *Appl. Phys. Lett.* **101**, 121910, (2012).

8 Gu, D. D., Meiners, W., Wissenbach, K. & Poprawe, R. Laser additive manufacturing of metallic components: materials, processes and mechanisms. *Int. Mater. Rev.* **57**, 133-164, (2012).

9 Yang, T. *et al.* Multicomponent intermetallic nanoparticles and superb mechanical behaviors of complex alloys. *Science* **362**, 933-937, (2018).

10 Du, X. H. *et al.* Dual heterogeneous structures lead to ultrahigh strength and uniform ductility in a Co-Cr-Ni medium-entropy alloy. *Nat. Commun.* **11**, 2390, (2020).

11 Gwalani, B. *et al.* Cu assisted stabilization and nucleation of L12 precipitates in Al0. 3CuFeCrNi2 fcc-based high entropy alloy. *Acta Mater.* **129**, 170-182, (2017).

12 Radecka, A. *et al.* The formation of ordered clusters in Ti–7Al and Ti–6Al–4V. *Acta Mater.* **112**, 141-149, (2016).

13 Zhang, R. *et al.* Direct imaging of short-range order and its impact on deformation in Ti-6Al. *Sci. Adv.* **5**, eaax2799, (2019).

14 Marceau, R. K. W., Ceguerra, A. V., Breen, A. J., Raabe, D. & Ringer, S. P. Quantitative chemical-structure evaluation using atom probe tomography: Short-range order analysis of Fe–Al. *Ultramicroscopy* **157**, 12-20, (2015).

15 Gault, B., Moody, M. P., Cairney, J. M. & Ringer, S. P. *Atom probe microscopy*. Vol. 160 (Springer, 2012).

16 Geiser, B. P., Kelly, T. F., Larson, D. J., Schneir, J. & Roberts, J. P. Spatial Distribution Maps for Atom Probe Tomography. *Microsc. Microanal.* **13**, 437-447, (2007).

17 Gault, B. *et al.* Atom probe microscopy investigation of Mg site occupancy within δ′ precipitates in an Al–Mg–Li alloy. *Scripta Mater.* **66**, 903-906, (2012).

18 Meher, S. & Banerjee, R. Partitioning and site occupancy of Ta and Mo in Co-base $\gamma/\gamma'$ alloys studied by atom probe tomography. *Intermetallics* **49**, 138-142, (2014).

19 Garcia-Gonzalez, M. *et al.* Influence of thermo-mechanical history on the ordering kinetics in 18 carat Au alloys. *Acta Mater.* **191**, 186-197, (2020).

20 Zhang, F. X. *et al.* Local Structure and Short-Range Order in a NiCoCr Solid Solution Alloy. *Phys. Rev. Lett.* **118**, 205501, (2017).

21 Ding, Q. *et al.* Tuning element distribution, structure and properties by composition in high-entropy alloys. *Nature* **574**, 223-227, (2019).

22 Wei, Y. *et al.* Machine-learning-based atom probe crystallographic analysis. *Ultramicroscopy* **194**, 15-24, (2018).

23 Zelenty, J., Dahl, A., Hyde, J., Smith, G. D. W. & Moody, M. P. Detecting Clusters in Atom Probe Data with Gaussian Mixture Models. *Microsc. Microanal.* **23**, 269-278, (2017).

24 Bapst, V. *et al.* Unveiling the predictive power of static structure in glassy systems. *Nature Physics* **16**, 448-454, (2020).

25 Azimi, S. M., Britz, D., Engstler, M., Fritz, M. & Mücklich, F. Advanced Steel Microstructural Classification by Deep Learning Methods. *Sci. Rep.* **8**, 2128, (2018).

26 LeCun, Y., Bottou, L., Bengio, Y. & Haffner, P. Gradient-based learning applied to document recognition. *Proc. IEEE* **86**, 2278-2324, (1998).

27 Krizhevsky, A., Sutskever, I. & Hinton, G. E. Imagenet classification with deep convolutional neural networks. *Commun. ACM* **60**, 84-90, (2017).

28 Williams, C. A., Haley, D., Marquis, E. A., Smith, G. D. & Moody, M. P. Defining clusters in APT reconstructions of ODS steels. *Ultramicroscopy* **132**, 271-278, (2013).

29 Du, Q. *et al.* Modeling over-ageing in Al-Mg-Si alloys by a multi-phase CALPHAD-coupled Kampmann-Wagner Numerical model. *Acta Mater.* **122**, 178-186, (2017).

30 Li, Y. *et al.* Precipitation and strengthening modeling for disk-shaped particles in aluminum alloys: size distribution considered. *Materialia* **4**, 431-443, (2018).

31 Immerkaer, J. Fast noise variance estimation. *Computer vision and image understanding* **64**, 300-302, (1996).

32 Bishop, C. M. Training with noise is equivalent to Tikhonov regularization. *Neural computation* **7**, 108-116, (1995).

33 Saito, T. & Rehmsmeier, M. Precrec: fast and accurate precision–recall and ROC curve calculations in R. *Bioinformatics* **33**, 145-147, (2017).

34 Kannan, R. & Vasanthi, V. in *Soft Computing and Medical Bioinformatics* 63-72 (Springer, 2019).

35 Chollet, F. *Deep Learning mit Python und Keras: Das Praxis-Handbuch vom Entwickler der Keras-Bibliothek.* (MITP-Verlags GmbH & Co. KG, 2018).

36 Liu, R., Agrawal, A., Liao, W.-k., Choudhary, A. & De Graef, M. Materials discovery: Understanding polycrystals from large-scale electron patterns. *2016 IEEE International Conference on Big Data (Big Data)*, 2261-2269, (2016).

37 Selvaraju, R. R. *et al.* Grad-cam: Visual explanations from deep networks via gradient-based localization. *Proceedings of the IEEE international conference on computer vision*, 618-626, (2017).

38 Kotikalapudi, R. & Contributors. *keras-vis*, <https://github.com/raghakot/keras-vis> (2017).

39 Ding, J., Yu, Q., Asta, M. & Ritchie, R. O. Tunable stacking fault energies by tailoring local chemical order in CrCoNi medium-entropy alloys. *Proc. Natl. Acad. Sci. U.S.A.* **115**, 8919-8924, (2018).

40 Tan, C. *et al.* A survey on deep transfer learning. *International conference on artificial neural networks*, 270-279, (2018).

41 Lakshminarayanan, B., Pritzel, A. & Blundell, C. Simple and scalable predictive uncertainty estimation using deep ensembles. *Advances in neural information processing systems*, 6402-6413, (2017).

42 Gal, Y. & Ghahramani, Z. Dropout as a bayesian approximation: Representing model uncertainty in deep learning. *international conference on machine learning*, 1050-1059, (2016).

43 Wenzel, F. *et al.* How good is the bayes posterior in deep neural networks really? *arXiv preprint arXiv:2002.02405*, (2020).

44 Gault, B. *et al.* Estimation of the Reconstruction Parameters for Atom Probe Tomography. *Microsc. Microanal.* **14**, 296-305, (2008).

45 Gault, B. *et al.* Advances in the calibration of atom probe tomographic reconstruction. *J. Appl. Phys.* **105**, 034913, (2009).

46 Goodfellow, I., Bengio, Y. & Courville, A. *Deep learning*. (MIT press, 2016).

47 Tieleman, T. & Hinton, G. Lecture 6.5-rmsprop: Divide the gradient by a running average of its recent magnitude. *COURSERA: Neural networks for machine learning* **4**, 26-31, (2012).

48 Chollet, F. keras. GitHub repository. *https://github. com/fchollet/keras*, (2015).

49 Abadi, M. *et al.* TensorFlow: Large-scale machine learning on heterogeneous systems. (2015).

## Table

**Table 1** Parameters for generating the crystal structure library and generating the corresponding *zx*-SDMs dictionary. Note that "σ" represented the standard deviation of the Gaussian function.

| Category | Number of *zx*-SDMs | $\sigma_x = \sigma_y$, nm | $\sigma_z$, nm | Detect efficiency | Rotation, degree |
|---|---|---|---|---|---|
| FCC structure | 9866 | 0.2~0.8 | 0.01~0.04 | 0.35~0.8 | -1~1 |
| $L1_2$ structure | 8550 | 0.2~0.8 | 0.03~0.055 | 0.35~0.8 | -1~1 |

**Figure legends**

Figure 1 **APT data of Al-5.18Mg-6.79Li (at.%) alloy.** (a) 3D atom probe tomography; (b) 2D desorption map histogram; (c) reconstructed 3D atom probe tomography along [$\bar{1}01$] crystallographic pole; (d) a close-up of a thin slice in (c); examples of *zx*-SDMs along [$\bar{1}01$]: Al-Al interatomic vector tomograms of (e) the FCC matrix and (f) a $L1_2$ precipitate. The corresponding unit cell models are also given.

Figure 2 **Procedure to build crystal structure library and generate simulated *zx*-SDMs dictionary.** (a) Generating supercell; (b) making Euler transformation; (c) adding Gaussian noise to shift atoms in x, y, z directions; (d) removing part of atoms; (e) simulated *zx*-SDMs of Al-Al pairs. Note that the Mg and Li atoms are not separately labelled in the $L1_2$ structure.

Figure 3 **Proposed CNN configuration.** (a) The architecture of CNN for identifying ordered structures; (b) the detailed architecture of each layer; (c) the final loss values of training and validation after 7 epochs of different CNN structures including different convolutional (plus max pooling) layers as shown in (a) only using simulated data.

Figure 4 **Test results in experimental data.** Predicted $L1_2$ structure probability of a small volume of APT data (Figure 1 (c)) in (a) case 1 and (b) case 2. The corresponding part *zx*-SDMs are attached. (c) ROC curves of the two cases.

Figure 5 **Prediction of the CNN-based $L1_2$-ordered structure recognition model in Al-Li-Mg APT data.** (a) Frequency distribution of the predicted $L1_2$ structure probabilities of the 1-nm

voxels generated from the dataset shown in Figure 1 (c) based on Supplementary Figure 3; (b) the *zx*-SDMs generated from the data corresponding to zone 1, 2, and 3 in (a).

Figure 6 **Comparison of 3D atom tomography of $L1_2$ ordered structures revealed by two methods.** 3D atom probe tomography of precipitates (a) by the proposed CNN approach with the predicted $L1_2$ structure probability (P) above 62 and (d) by the previous report using 8 at.% Li isocomposition. For (a), the corresponding species-specific intensity distributions along the z-axis in the *z*-SDMs with the $L1_2$ structure probability above (b) and below (c) 62. For (d), the corresponding species-specific intensity distributions along the z-axis in the *z*-SDMs of above (e) and below (f) 8 at.% Li.

Figure 7 **Visualization of the CNN model.** Part of (a) feature maps and (b) Grad-CAM of *zx*-SDMs of the simulated FCC and $L1_2$ structures after the first to the fourth convolutional layers in the applied CNN.

Figure 1

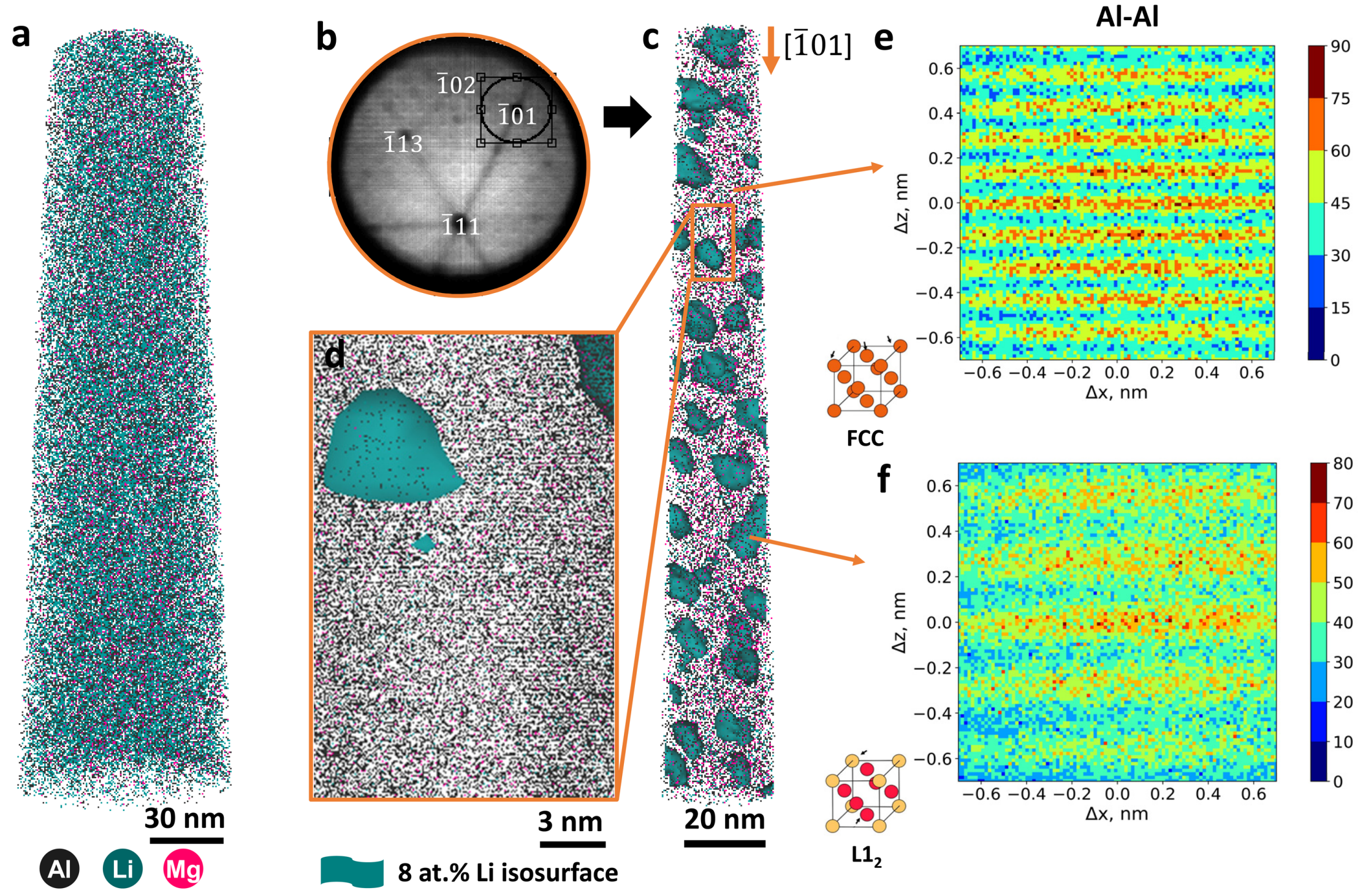

Figure 2

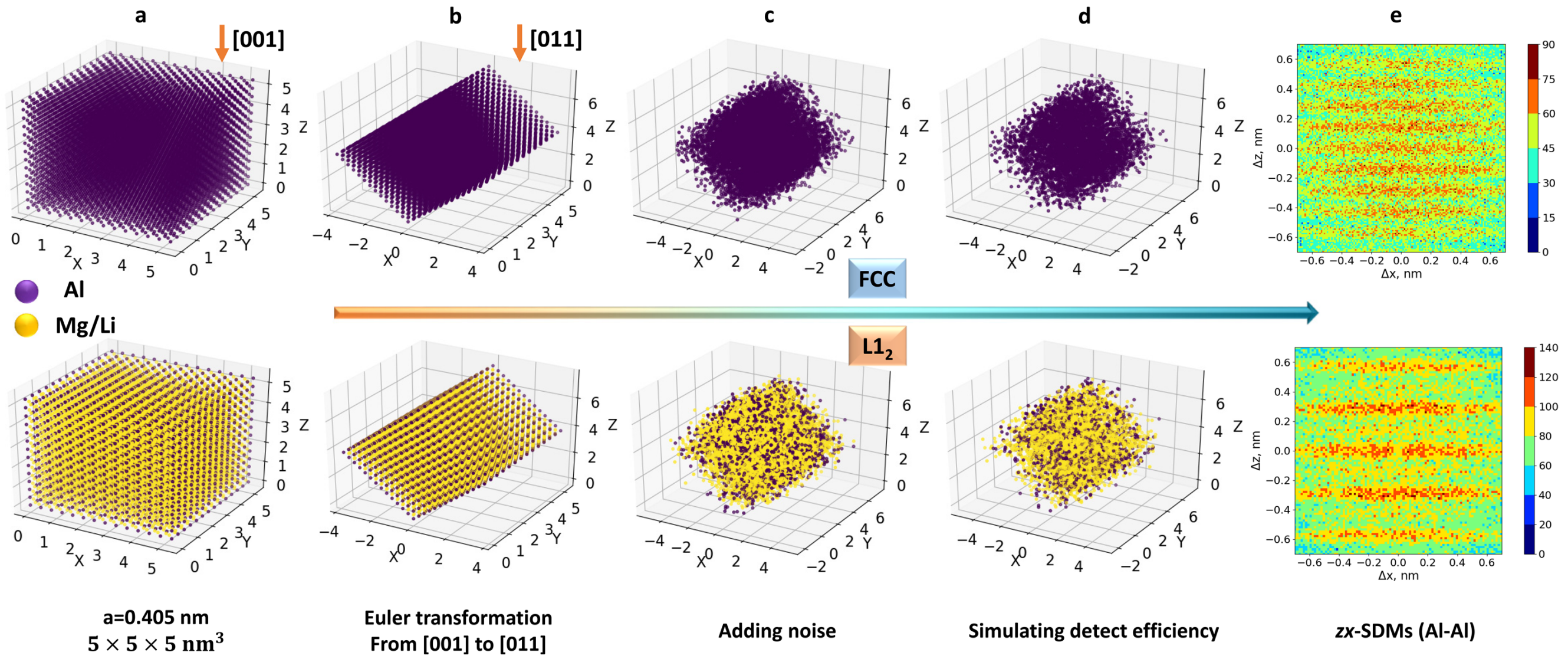

Figure 3

**a**

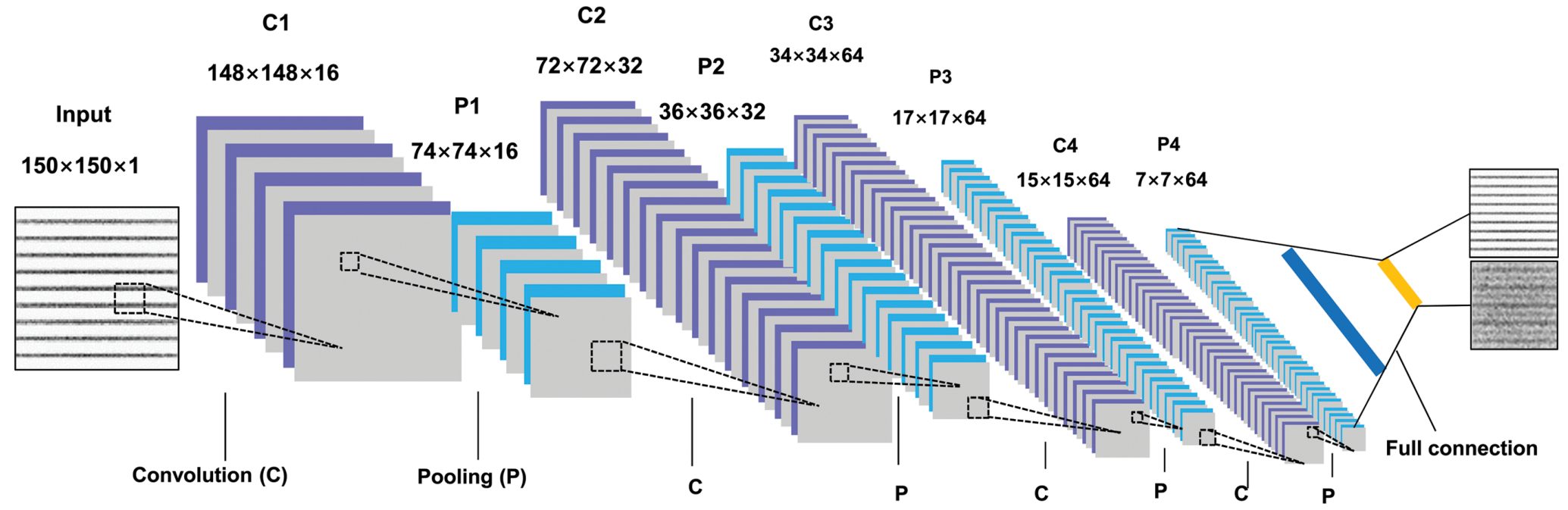

**b** Architecture of the used convolutional neural network

| Layer type | Specifications |
|---|---|
| Convolutional layer 1 | Kernel: 3 × 3; 16 filters |
| Max pooling layer 1 | Pool size: 2 × 2; stride: 2 × 2; padding='valid' |
| Convolutional layer 2 | Kernel: 3 × 3; 32 filters |
| Max pooling layer 2 | Pool size: 2 × 2; stride: 2 × 2; padding='valid' |
| Convolutional layer 3 | Kernel: 3 × 3; 64 filters |
| Max pooling layer 3 | Pool size: 2 × 2; stride: 2 × 2; padding='valid' |
| Convolutional layer 4 | Kernel: 3 × 3; 64 filters |
| Max pooling layer 4 | Pool size: 2 × 2; stride: 2 × 2; padding='valid' |
| Fully connected layer | Size: 512 |
| Output layer | Size: 2 |

**c**

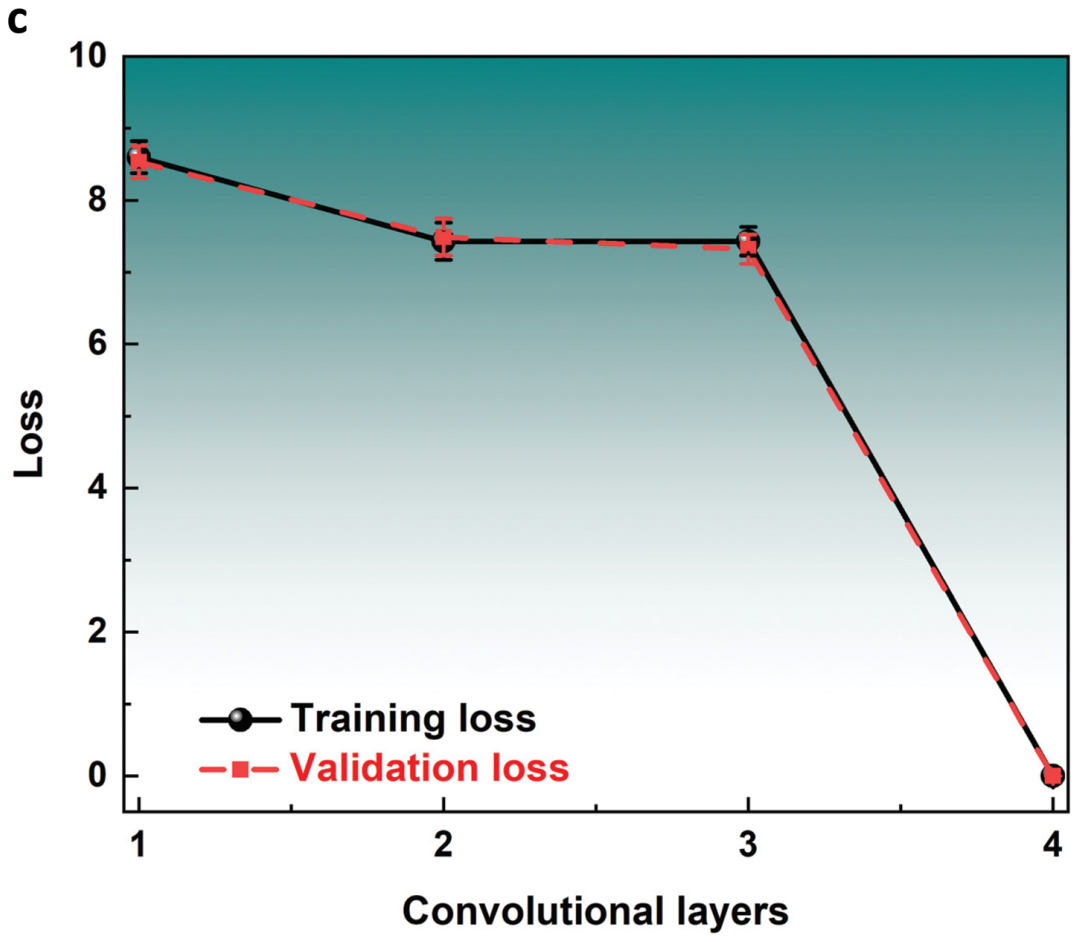

Figure 4

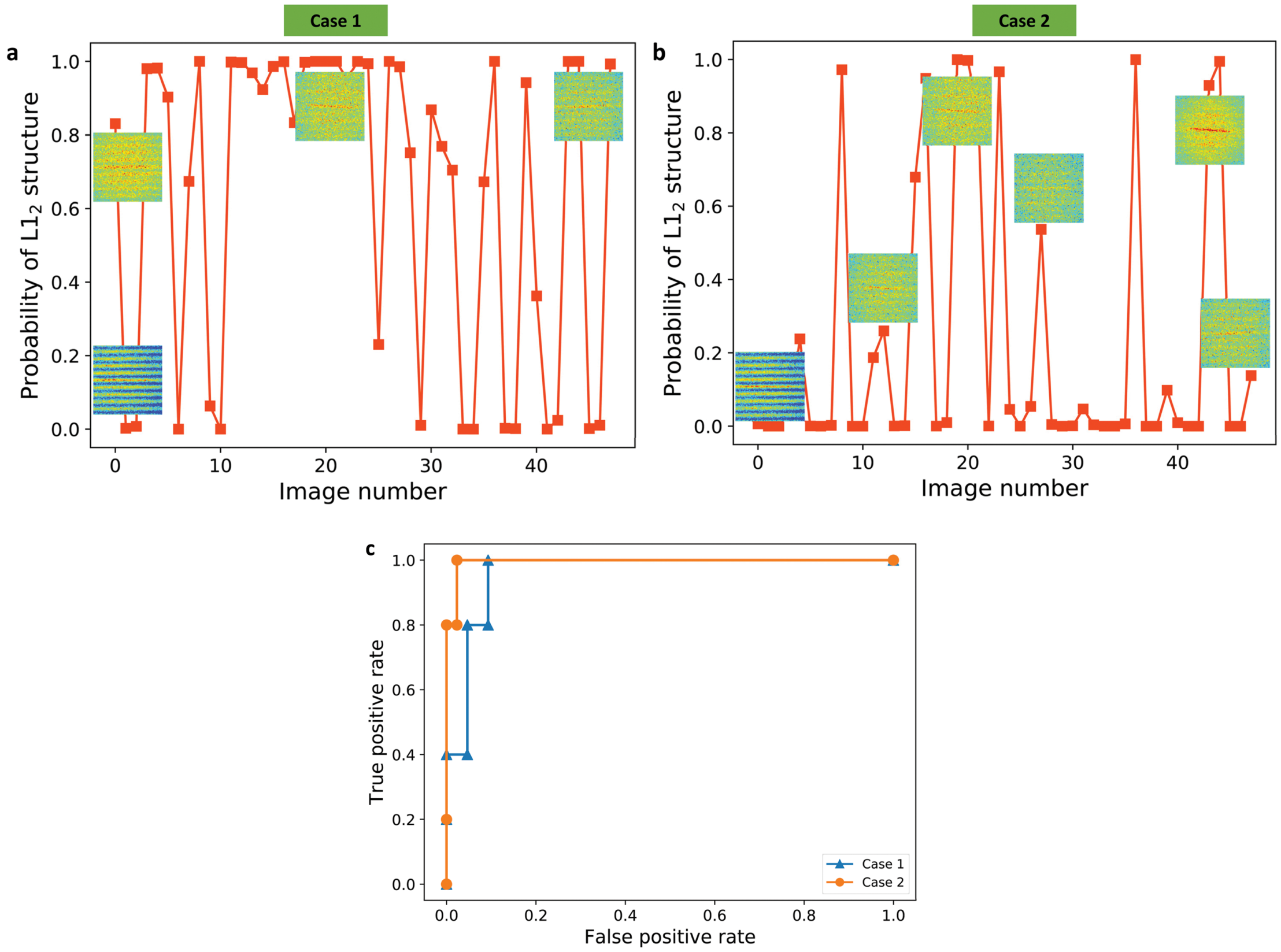

Figure 5

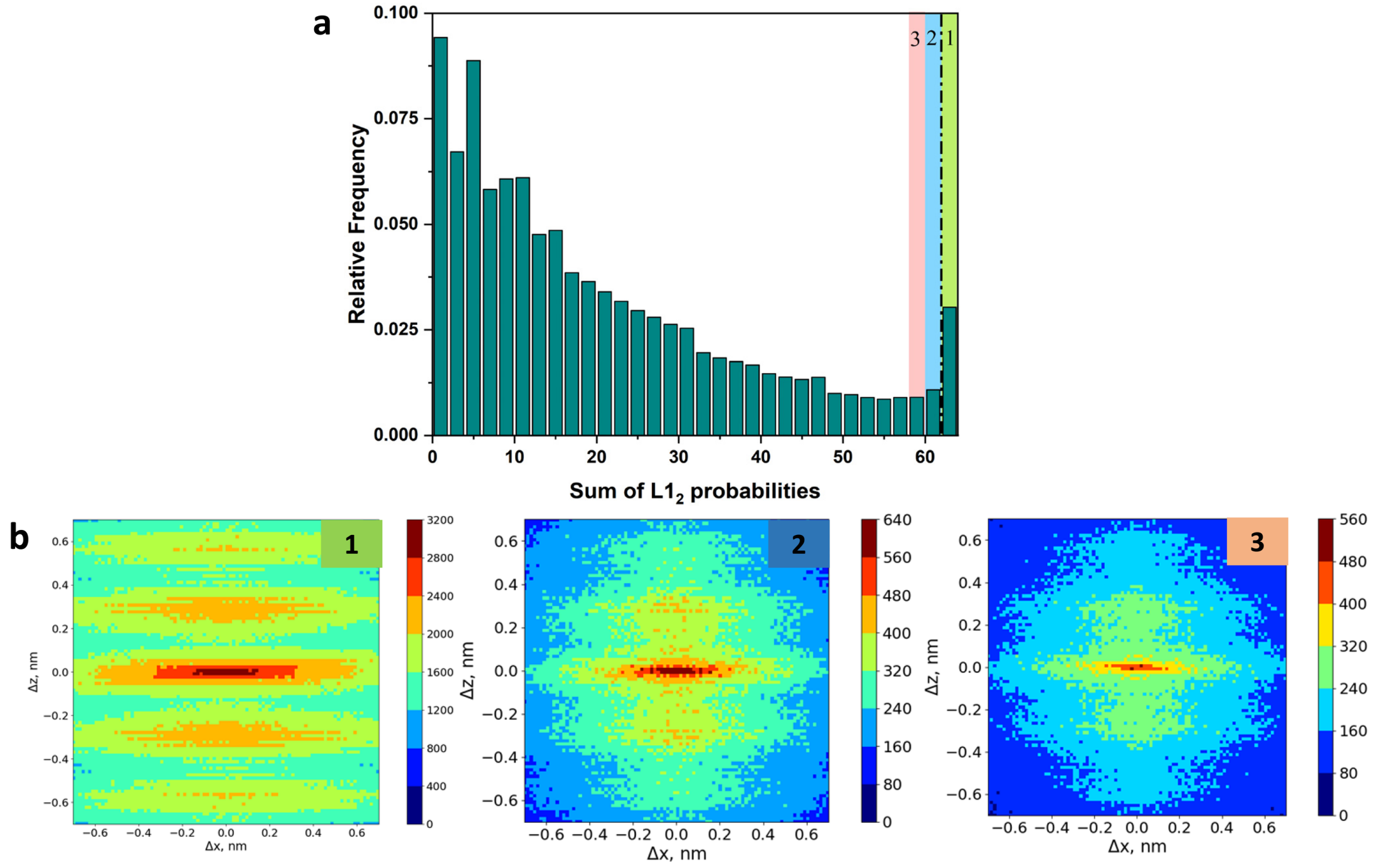

Figure 6

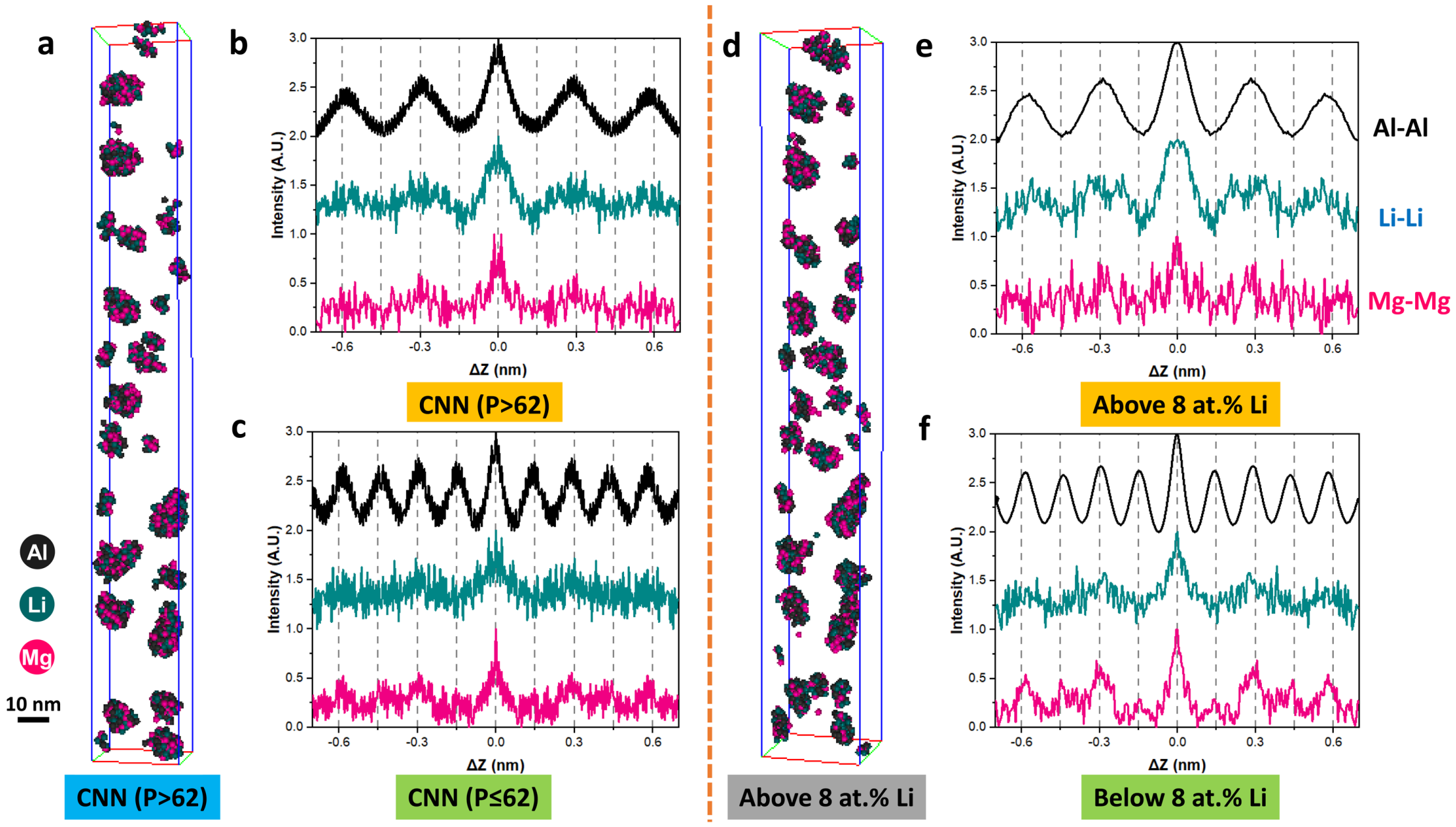

Fig 7

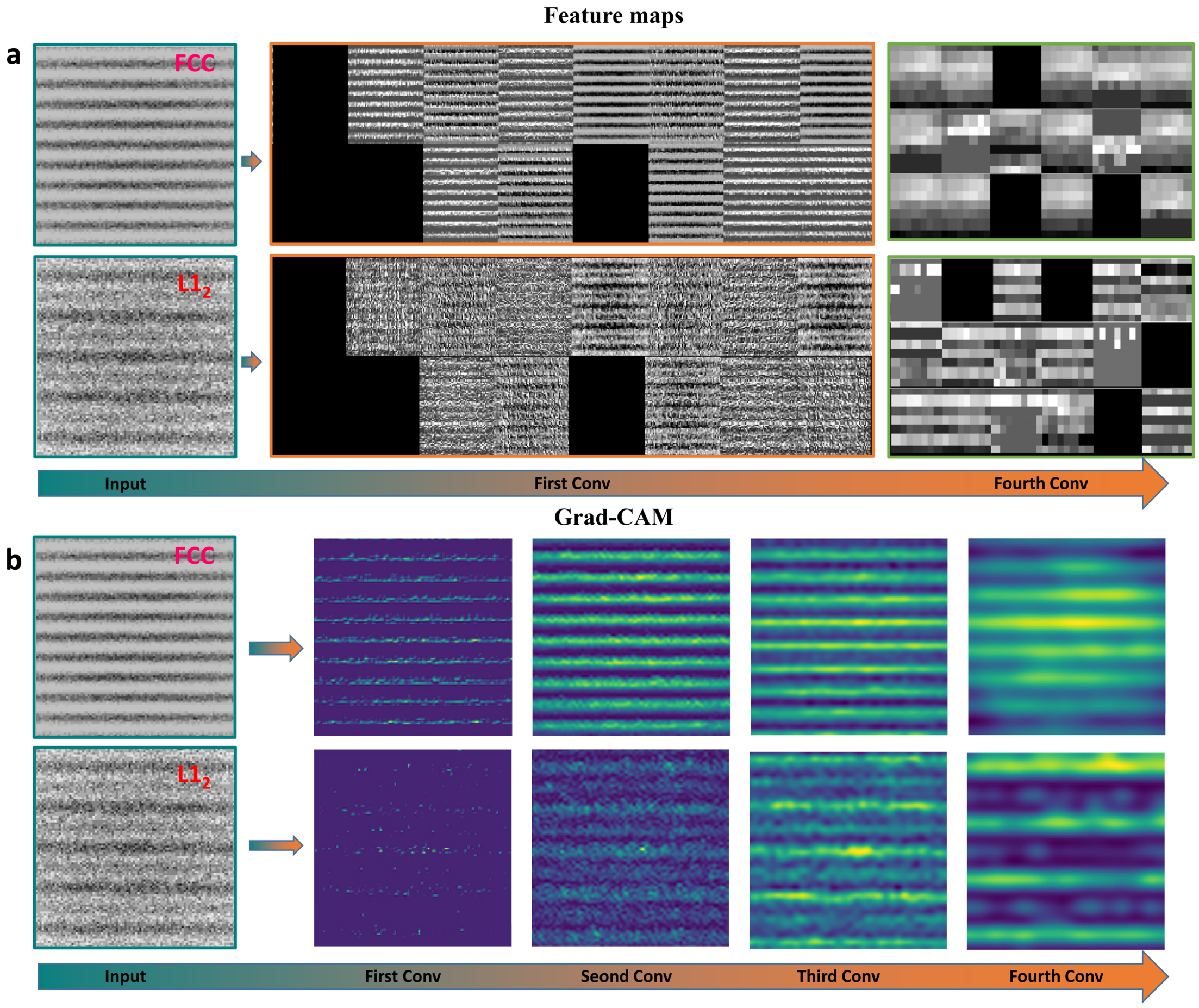

**Supplementary information for**

# Convolutional neural network-assisted recognition of nanoscale $L1_2$ ordered structures in face-centred cubic alloys

Yue Li[1,*], Xuyang Zhou[1], Timoteo Colnaghi[2], Ye Wei[1],

Andreas Marek[2], Hongxiang Li[3], Stefan Bauer[4], Markus Rampp[2], Leigh T. Stephenson[1,*]

[1] Max-Planck Institut für Eisenforschung GmbH, Max-Planck-Straße 1, 40237 Düsseldorf, Germany

[2] Max Planck Computing and Data Facility, Gießenbachstraße 2, 85748 Garching, Germany

[3] State Key Laboratory for Advanced Metals and Materials, University of Science and Technology Beijing, 100083, Beijing, China

[4] Max Planck Institute for Intelligent Systems, Max-Planck-Ring 4, 72076 Tübingen, Germany

[*]Corresponding authors, Yue Li, Tel.: +49 211 6792 871, E-mail: yue.li@mpie.de

Leigh T. Stephenson, Tel.: +49 211 6792 794, E-mail: l.stephenson@mpie.de

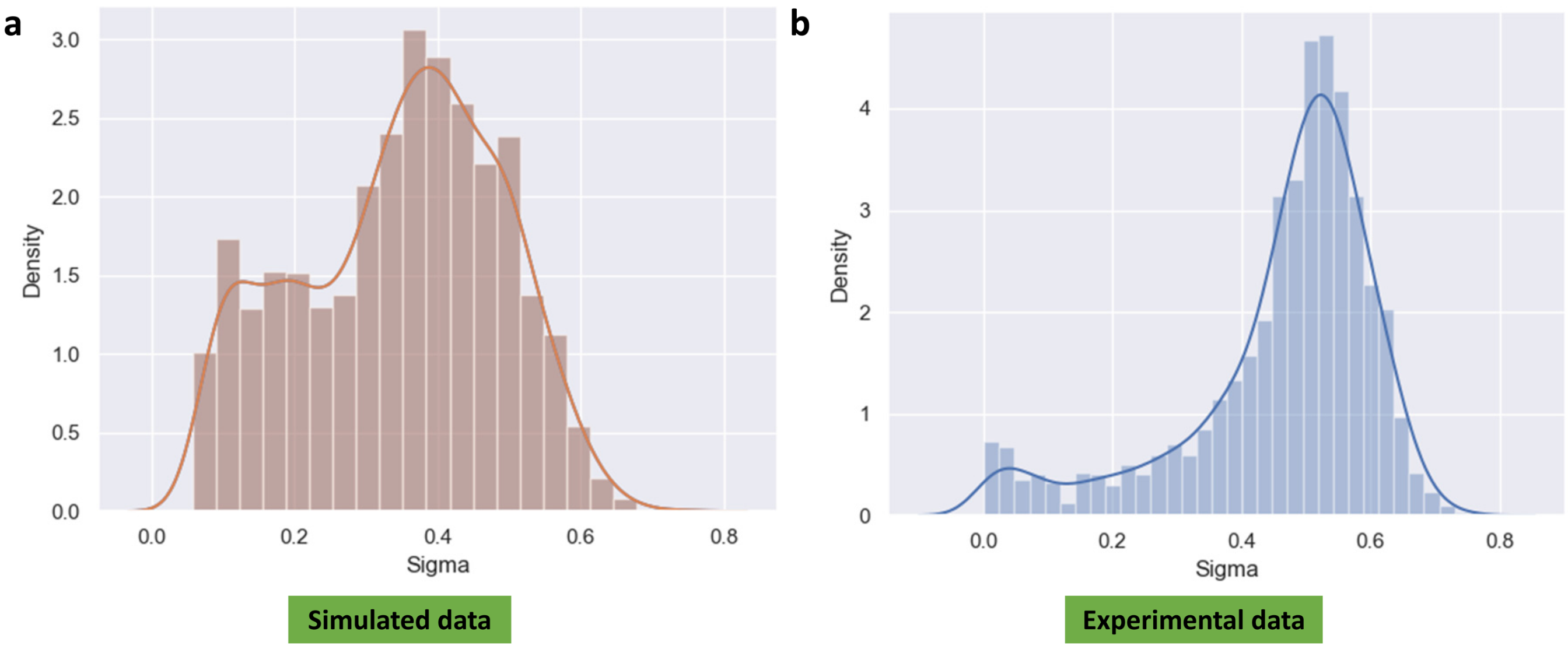

**Supplementary Figure 1:** Noise variance (Sigma$^2$) estimation of the randomly-chosen (a) simulated (~2000 images) and (b) experimental (~2000 images) data using zero mean Gaussian noise. The corresponding Gaussian kernel density estimate curves are given.

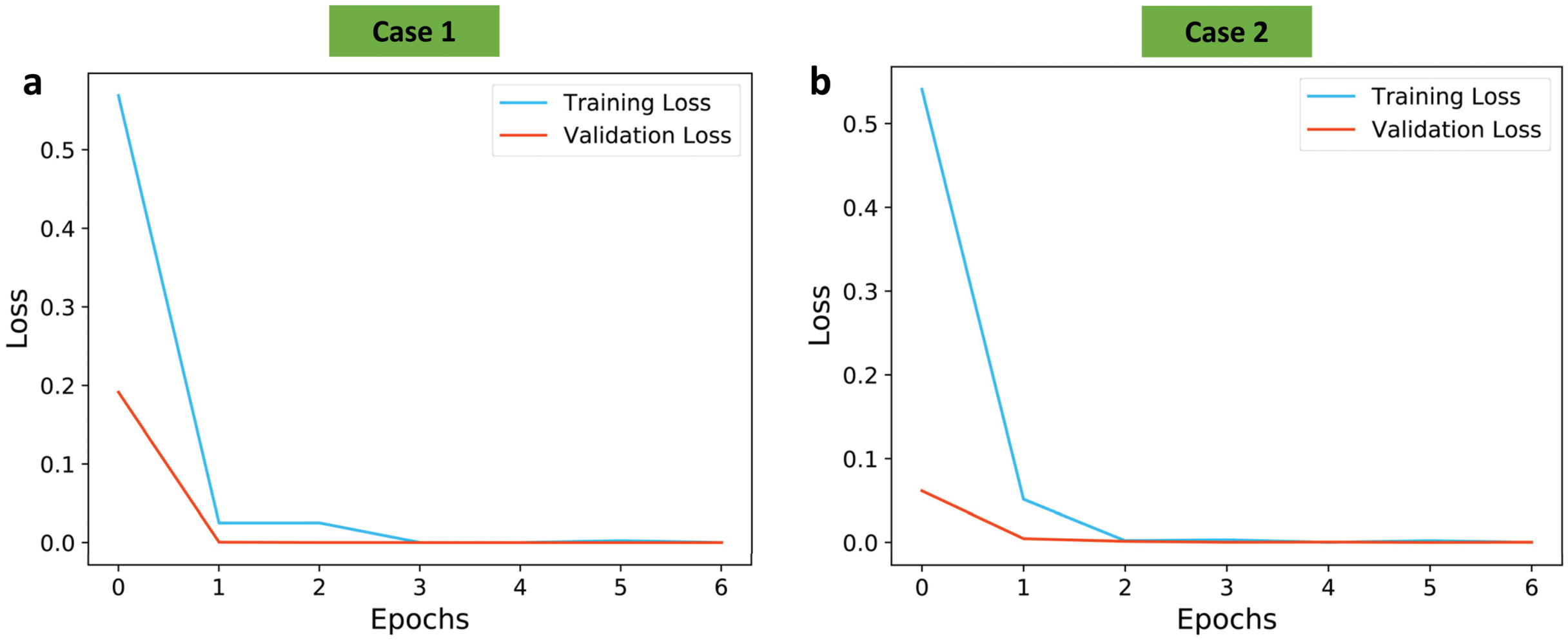

**Supplementary Figure 2:** History of training and validation losses of the optimum CNN model after fivefold cross-validation in (a) case 1 and (b) case 2.

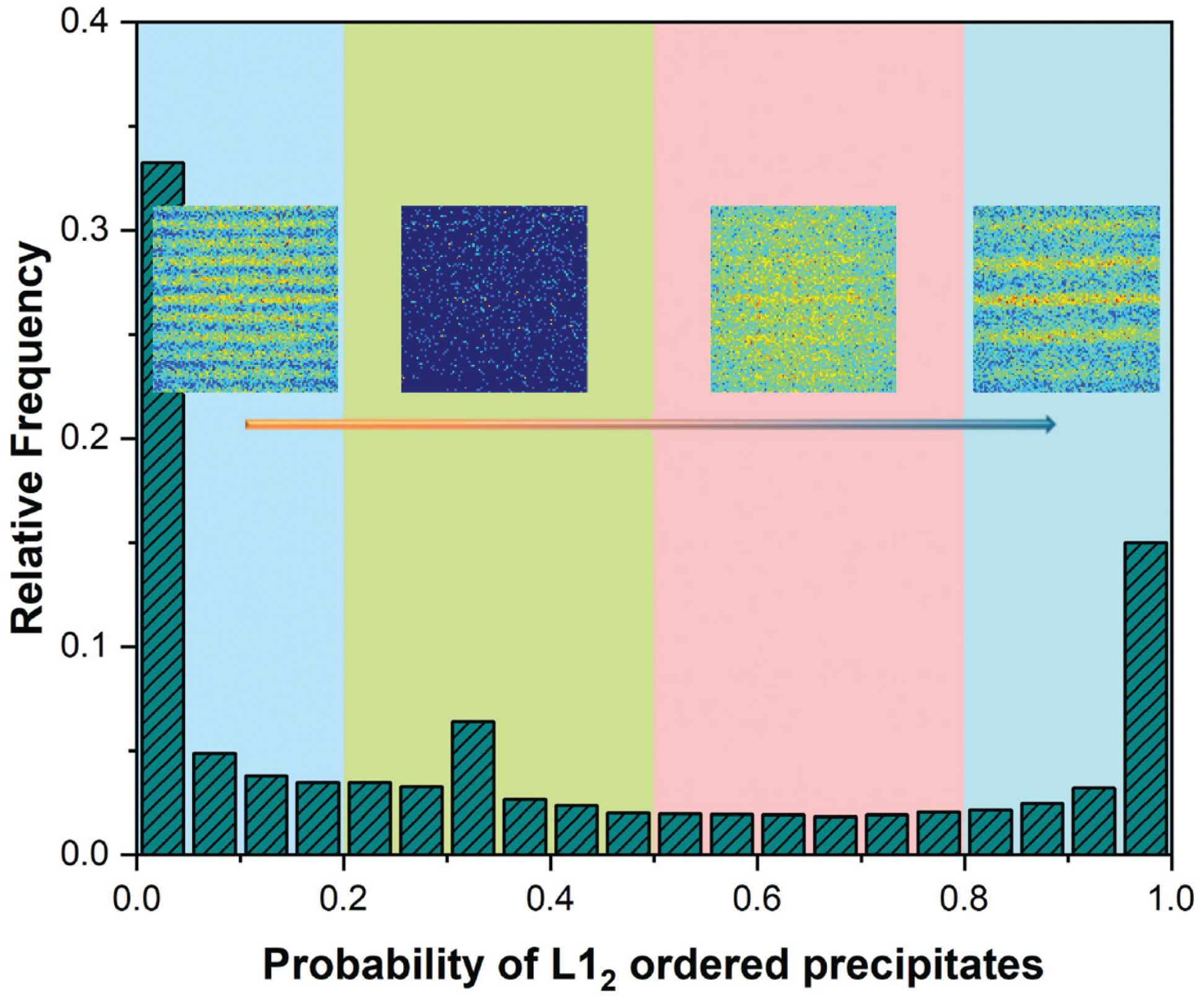

**Supplementary Figure 3:** Frequency distribution of the predicted $L1_2$ structure probabilities of 98175 4-nm voxels generated from the dataset shown in Figure 1 (c). Four *zx*-SDMs corresponding to different probability values are attached. Note that the peak of about 0.3 results from *zx*-SDMs generated from limited data in the 4-nm cubes.

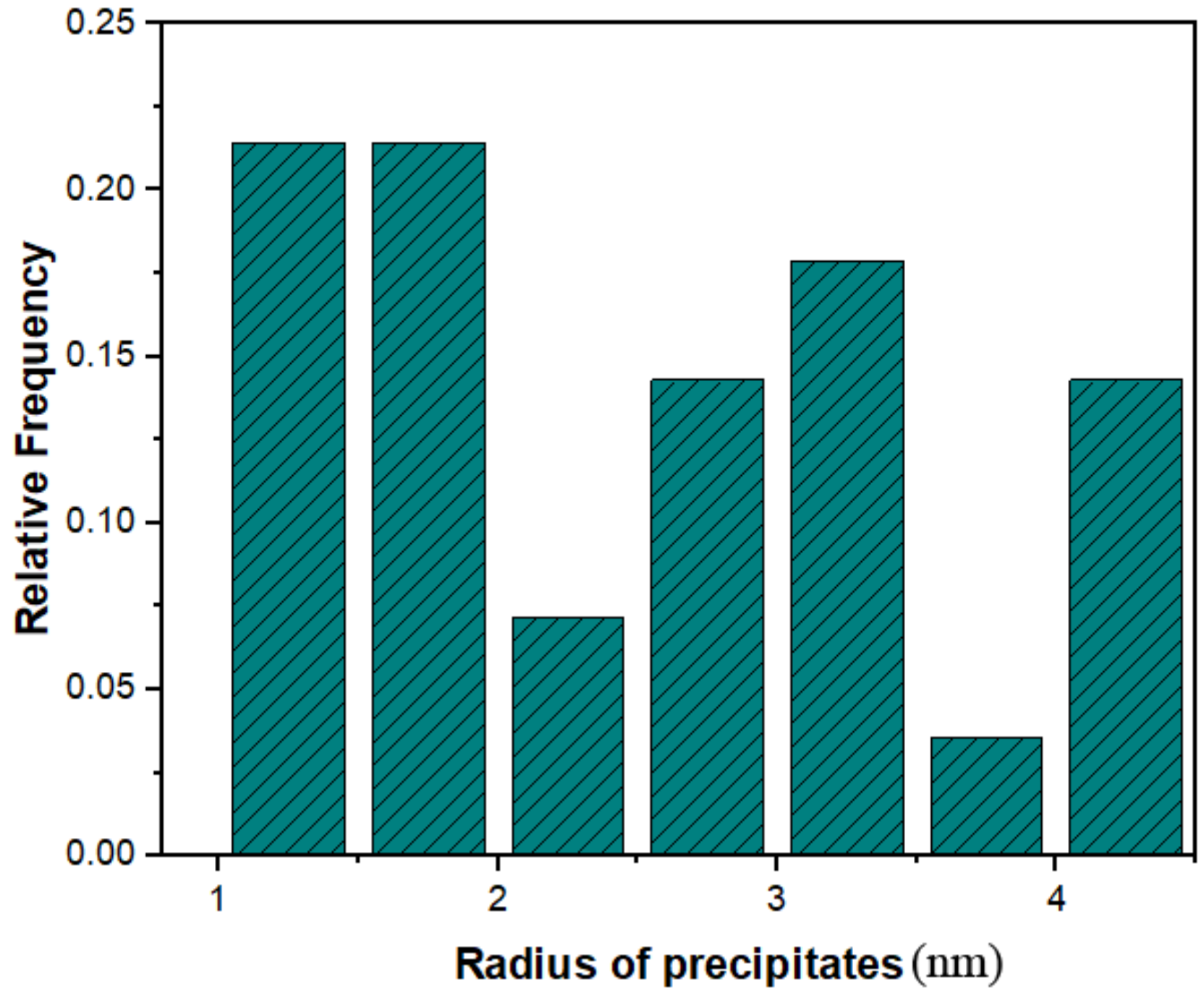

**Supplementary Figure 4:** Precipitate size distribution of $L1_2$ structure obtained from APT data (Figure 1 (c)) based on the proposed CNN-APT method.

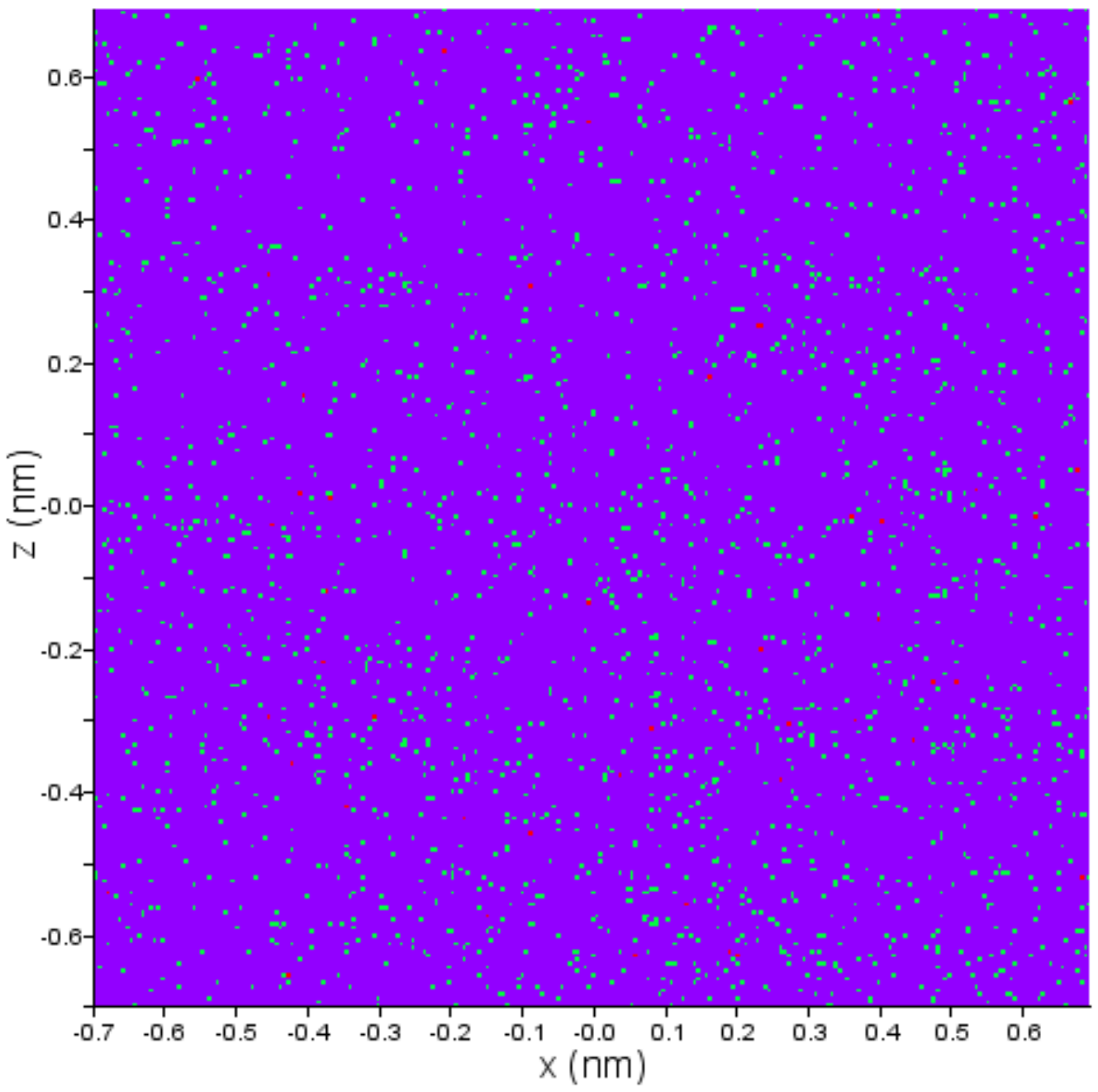

(a)

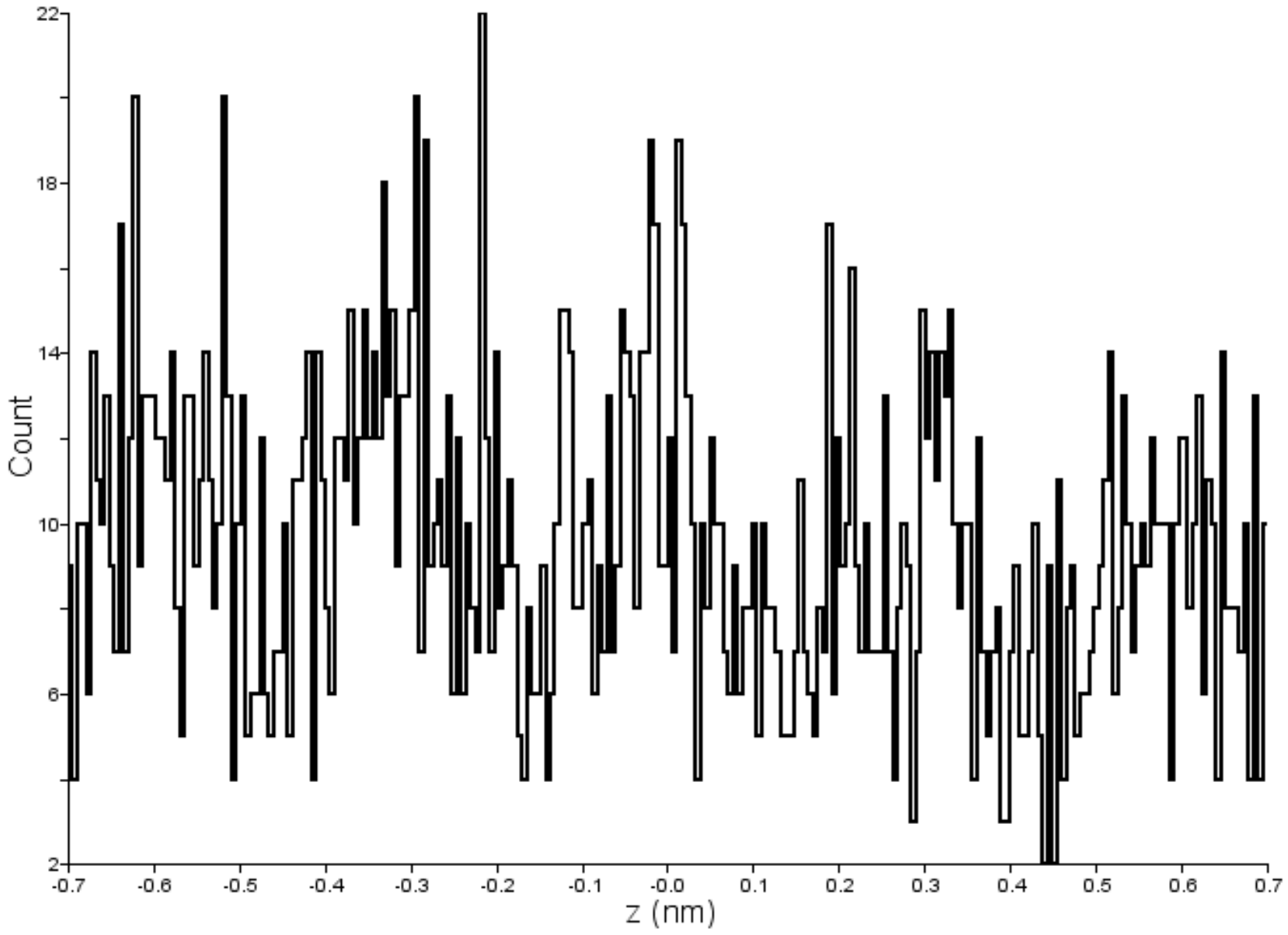

(b)

**Supplementary Figure 5:** An example of (a) 2D *zx*-SDM and (b) 1D Al-Al *z*-SDM of a 1-nm voxel extracted from Figure 6 (a).